\documentclass[aps,pra,reprint,amsmath,amssymb,superscriptaddress,nobibnotes]{revtex4-1}
\usepackage{graphicx,SIunits}
\usepackage{bm}     % bold math
\usepackage{hyperref} % add hypertext capabitlities
\usepackage{dsfont}    % allows \mathds{}
\usepackage{color}

\begin{document}

\title{Characterizing Bell state analyzer using weak coherent pulses}

\author{Donghwa Lee}
\affiliation{Center for Quantum Information, Korea Institute of Science and Technology (KIST), Seoul, 02792, Republic of Korea}
\affiliation{Division of Nano \& Information Technology, KIST School, Korea University of Science and Technology, Seoul 02792, Republic of Korea}

%\author{Seong-Jin Hong}
%\affiliation{Center for Quantum Information, Korea Institute of Science and Technology (KIST), Seoul, 02792, Republic of Korea}

\author{Young-Wook Cho}
\affiliation{Center for Quantum Information, Korea Institute of Science and Technology (KIST), Seoul, 02792, Republic of Korea}

\author{Hyang-Tag Lim}
\affiliation{Center for Quantum Information, Korea Institute of Science and Technology (KIST), Seoul, 02792, Republic of Korea}

\author{Sang-Wook Han}
\affiliation{Center for Quantum Information, Korea Institute of Science and Technology (KIST), Seoul, 02792, Republic of Korea}
\affiliation{Division of Nano \& Information Technology, KIST School, Korea University of Science and Technology, Seoul 02792, Republic of Korea}

\author{Hojoong Jung}
\affiliation{Center for Quantum Information, Korea Institute of Science and Technology (KIST), Seoul, 02792, Republic of Korea}

\author{Sung Moon}
\affiliation{Center for Quantum Information, Korea Institute of Science and Technology (KIST), Seoul, 02792, Republic of Korea}
\affiliation{Division of Nano \& Information Technology, KIST School, Korea University of Science and Technology, Seoul 02792, Republic of Korea}

%\author{Kwangjo Lee}
%\affiliation{Department of Applied Physics, Kyung Hee University, Yongin, 17104, Republic of Korea}

\author{Yong-Su Kim}
\email{yong-su.kim@kist.re.kr}
\affiliation{Center for Quantum Information, Korea Institute of Science and Technology (KIST), Seoul, 02792, Republic of Korea}
\affiliation{Division of Nano \& Information Technology, KIST School, Korea University of Science and Technology, Seoul 02792, Republic of Korea}

\date{\today} 

\begin{abstract}
\noindent Bell state analyzer (BSA) is one of the most crucial apparatus in photonic quantum information processing. While linear optics provide a practical way to implement BSA, it provides unavoidable errors when inputs are not ideal single-photon states. Here, we propose a simple method to deduce the BSA for single-photon inputs using weak coherent pulses. By applying the method to Reference-Frame-Independent Measurement-Device-Independent Quantum Key Distribution, we experimentally verify the feasibility and effectiveness of the method.
\end{abstract}

\keywords{Multiparty quantum communication, Entanglement, Information symmetry}

\maketitle

%\section{Introduction}

\section{Introduction}

%\noindent {\it Introduction.--} 

Entanglement is at the heart of quantum information processing~\cite{horodecki08}. Bell state analyzer (BSA), an experimental apparatus which performs projective measurement onto maximally entangled two-qubit states, plays central roles in many photonic quantum information processing such as fundamental quantum physics~\cite{branciard13,kim18}, quantum key distribution~\cite{braunstein12,lo12}, quantum teleportation~\cite{bennett93,bouwmeester97,gottesman99}, and quantum computation~\cite{knill01,gao10}. %In photonic qubit system, an ideal implementation of BSA is not straightforward since it requires photon-photon interaction~\cite{horodecki08}. Instead, the linear optical BSA based on two-photon interference and post-selection has been widely applied for various quantum information processing due to the simplicity and robustness of the implementation~\cite{xx} The linear optical BSA, however, has a few drawbacks comparing to the ideal one. 

\begin{figure}[b]
\centering\includegraphics[width=3.4in]{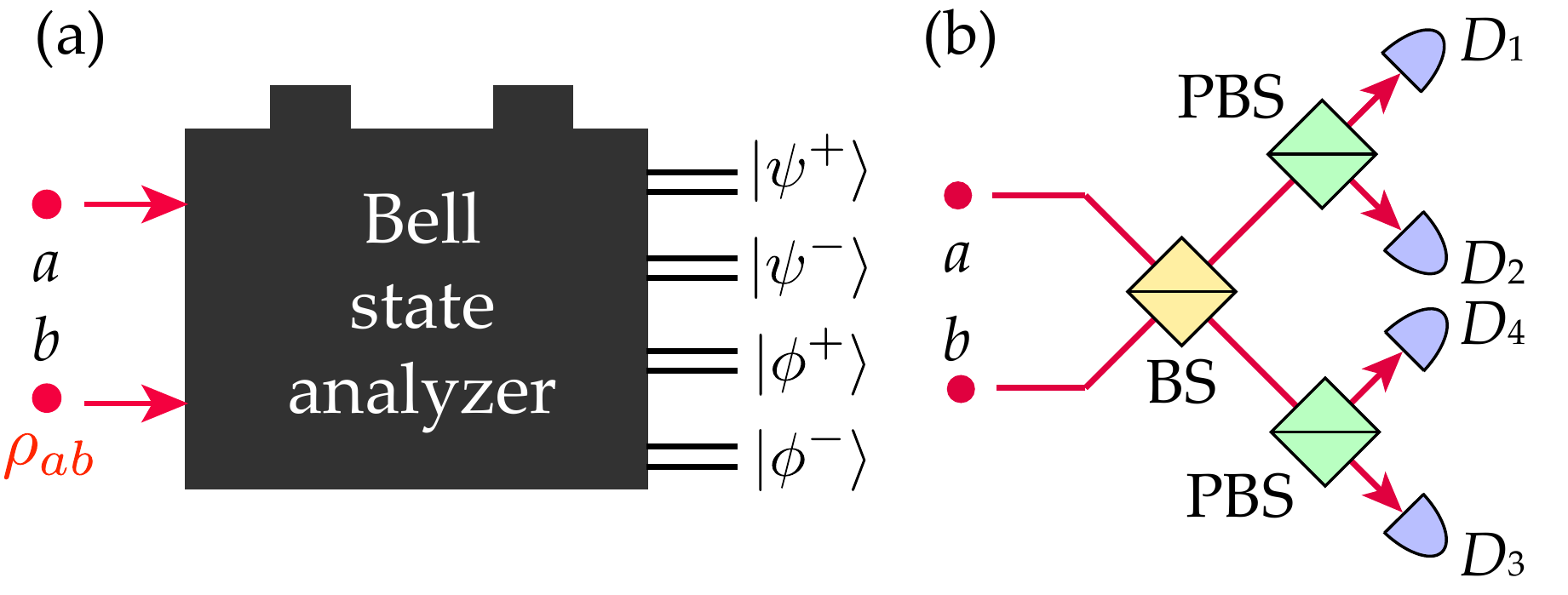}
\caption{(a) Conceptual diagram of an ideal Bell state analyzer. Any two-qubit input state $\rho_{ab}$ returns one of the four Bell states, $|\psi^{\pm}\rangle_{ab}$ and $|\phi^{\pm}\rangle_{ab}$. If the input is not a two-qubit state, it returns a $null$ outcome. (b) Polarization qubit Bell state analyzer using linear optics. BS : Beamsplitter, PBS : Polarizing beamsplitter, D : Single-photon detector. Only two Bell states out of four can be determined using this scheme, so the intrinsic success probability is $p=1/2$. It can provide certain outcomes even if the input is not a two-qubit state.}
\label{bsa}
\end{figure}

Figure~\ref{bsa}(a) presents the conceptual diagram of an ideal BSA. When a two-qubit state $\rho_{ab}$ is given at the input modes $a$ and $b$, it returns one of the four Bell states with the probability of $p_i=\langle\Psi_i|\rho_{ab}|\Psi_i\rangle$. Here, $|\Psi_i\rangle$ where $i\in\{1,2,3,4\}$ denotes one of four Bell states. Note that the overall BSA success probability $\Sigma_ip_i=1$ for any two-qubit input states. If the input state is not prepared in the form of $\rho_{ab}$, e.g., mode $a$ has two particles while mode $b$ is vacuum, it provides a $null$ outcome. % since $p=|\langle\Psi_i|\rho|\Psi_i\rangle|=0$.

In a photonic qubit system, the ideal implementation of BSA is not straightforward since it requires photon-photon interaction~\cite{horodecki08}. Instead, the linear optical BSA based on two-photon interference and post-selection has been widely applied for various quantum information processing due to the simplicity and robustness of the implementation~\cite{mattle96,ma12}. Figure~\ref{bsa} (b) shows a typical linear optical BSA setup in polarization qubits. 

%%%%%%%%%%%%%%%%%%%%%%%%%
%%%%%%%%%%%%%%%%%%%%%%%%%

The linear optical BSA, however, has a few drawbacks. With the scheme of Fig.~\ref{bsa} (b), $|\psi^\pm\rangle_{ab}=\frac{1}{\sqrt{2}}\left(|01\rangle\pm|10\rangle\right)=\frac{1}{\sqrt{2}}\left(|HV\rangle\pm|VH\rangle\right)$ input state is detected by $D_{12}$ or $D_{34}$ ($D_{14}$ or $D_{23}$) where $D_{ij}$ denotes the coincidence detection between $D_i$ and $D_j$. Here, $|H\rangle$ and $|V\rangle$ denote horizontal and vertical polarization states, respectively. The other two Bell states of $|\phi^\pm\rangle_{ab}=\frac{1}{\sqrt{2}}\left(|00\rangle\pm|11\rangle\right)$ cannot be detected by this scheme, so the intrinsic success probability is $p=1/2$. One can think of other configurations for BSA, however, it is known that the success probability of the linear optical BSA can be at most $p=1/2$~\cite{calsamiglia01}. Besides, the linear optical BSA works properly only when both input modes $a$ and $b$ are strictly occupied by single-photon states. In other words, it can provide incorrect outcomes when the input state is not prepared in $\rho_{ab}$. For example, if input mode $a$ has two photons and $b$ is vacuum, the linear optical BSA can tell one of the Bell state outcomes whereas the ideal one returns a $null$ outcome. Therefore, the linear optical BSA should be carefully utilized with ideal single-photon inputs, or it can provide unwanted errors. 

Despite these imperfections, the linear optical BSA has been widely applied with non-ideal single-photon inputs. For instance, Measurement-Device-Independent Quantum Key Distribution (MDI-QKD), where an untrusted third party performs Bell state measurement (BSM) onto optical pulses from two distant communication parties, is usually implemented with weak coherent pulses (WCP)~\cite{choi16,yin16,park18,liu19}. The use of non-ideal optical pulses causes non-zero intrinsic quantum bit error rate (QBER) in certain bases. Note that in Reference-Frame-Independent MDI-QKD (RFI-MDI-QKD), a more advanced QKD protocol which simultaneously provides high level of security and implementation practicality, these non-zero QBERs suggest complicated parameter estimation associated with the security~\cite{wang15,wang17,liu18}. 

In this paper, based on the recently proposed technique to characterize linear optical networks using non-ideal input states~\cite{yuan16,navarrete18,aragoneses18,zhang20}, we propose a method to characterize a linear optical BSA using WCP. In particular, we present a method to deduce the BSA result for ideal single-photon input states using the experimental data with WCP. We apply this method to MDI-QKD and RFI-MDI-QKD and show that the security associated parameter estimation can be more simple and experiment friendly. 

\section{Theory}

\subsection{Characterizing Bell state analyzer}

%Following the analysis in Ref.~\cite{navarrete18,aragoneses18}, 
%\noindent {\it Theory.--} 

Let us discuss how one can characterize the linear optical BSA of Fig~\ref{bsa} (b) for ideal single-photon inputs using the experimental data with WCP. Assuming that we input WCP with the same average photon number at the input modes $a$ and $b$, $\mu_a=\mu_b=\mu$, the coincidence counts between $D_i$ and $D_j$ is presented as
%%%%%%%%%%%%%%%%%%%%%%%%%
\begin{eqnarray}
&&N(D_{ij})^{\mu,\mu}=\kappa_i\kappa_j\mu^2e^{-2\mu}P(D_{ij}|1_P,1_Q)
\label{input}\\
&&+\kappa_i\kappa_j\frac{\mu^2}{2}e^{-\mu}\{P(D_{ij}|2_P,0)+P(D_{ij}|0,2_Q)\}+\mathcal{O}(\mu^r),\nonumber
\end{eqnarray}
%%%%%%%%%%%%%%%%%%%%%%%%%
where $P(D_{ij}|m_P,n_Q)$ is the conditional probability of coincidence detection $D_{ij}$ when the input mode $a~(b)$ is occupied by $m~(n)$ photons with $P~(Q)$ polarization state. Here, $\kappa_i$ denotes the detection efficiency of $D_i$. By blocking one of the input modes, one can obtain the coincidence counts as
%%%%%%%%%%%%%%%%%%%%%%%%%
\begin{eqnarray}
N(D_{ij})^{\mu,0}&=&\kappa_i\kappa_j\frac{\mu^2}{2}e^{-\mu}P(D_{ij}|2_P,0)+\mathcal{O}(\mu^r),\nonumber\\
N(D_{ij})^{0,\mu}&=&\kappa_i\kappa_j\frac{\mu^2}{2}e^{-\mu}P(D_{ij}|0,2_Q)+\mathcal{O}(\mu^r).
\label{input2}
\end{eqnarray}
%%%%%%%%%%%%%%%%%%%%%%%%%
Note that the higher order terms $\mathcal{O}(\mu^r)$ where $r\ge3$ is negligible for small average photon numbers $\mu\ll1$, so we will drop them for further analysis. Note that Eqs.~(\ref{input}) and~(\ref{input2}) ignore the cases when total number of input photons is smaller than two since they do not provide a coincidence detection.

From Eqs.~(\ref{input}) and (\ref{input2}), we can isolate $P(D_{ij}|1_P,1_Q)$ from all other $P(D_{ij}|m_P,n_Q)$ where $m_P\neq1$ and $n_Q\neq 1$ as
%%%%%%%%%%%%%%%%%%%%%%%%%
\begin{eqnarray}
P(D_{ij}|1_P,1_Q)=\frac{N(D_{ij})^{\mu,\mu}-N(D_{ij})^{\mu,0}-N(D_{ij})^{0,\mu}}{\kappa_i\kappa_j\mu^2e^{-2\mu}}.
\label{P_11}
\end{eqnarray}
%%%%%%%%%%%%%%%%%%%%%%%%%
While the numerator of Eq.~(\ref{P_11}) can be directly obtained from the experimental data, the denominator requires precise calibration of the detection efficiencies and the average photon number. In order to avoid the calibration problem, we investigate the sum of single-photon counts when one of the modes is blocked as
%%%%%%%%%%%%%%%%%%%%%%%%%
\begin{eqnarray}
N(D_i)=\kappa_i\mu e^{-\mu}\{P(D_i|1_P,0)+P(D_i|0,1_Q)\}, 
\label{D_i}
\end{eqnarray}
%%%%%%%%%%%%%%%%%%%%%%%%%
where $P(D_i|1_P,0)$ and $P(D_i|0,1_Q)$ are determined by the input polarization states. Therefore, one can characterize the behavior of linear optical BSA for ideal single-photon inputs using WCP with Eqs.~(\ref{P_11}) and (\ref{D_i}).

%In this case, Eq.~\eqref{P_11} becomes
%%%%%%%%%%%%%%%%%%%%%%%%%
%\begin{eqnarray}
%N(D_{ij}|1_P,1_Q)=\frac{P(D_{ij})^{\mu,\mu}-P(D_{ij})^{\mu,0}-P(D_{ij})^{0,\mu}}{\frac{1}{4}D_iD_j}.
%\label{P_11}
%\end{eqnarray}
%%%%%%%%%%%%%%%%%%%%%%%%%

%It is remarkable that $D_{13}$ and $D_{24}$, the non-zero coincidence detection events with single-photon inputs at one of the input mode $a$ and $b$, in the linear optical BSA does not provide a meaningful result. Therefore, 

\subsection{Application to RFI-MDI-QKD}

Let us first apply the above BSA characterization to MDI-QKD. In MDI-QKD, two communication parties, Alice and Bob,  transmit optical pulses to a third party who performs BSM~\cite{braunstein12,lo12}. Note that the MDI-QKD is usually implemented using WCP with decoy states~\cite{yin16,choi16,park18,liu19}. Let us first discuss the case when both communication parties transmit the states in $Z$-basis, i.e., either $|0\rangle=|H\rangle$ or $|1\rangle=|V\rangle$. Since a PBS transmits (reflects) the horizontal (vertical) polarization state, the horizontal and vertical polarization states are detected at $\{D_1,D_3\}$ and $\{D_2,D_4\}$, respectively. Therefore, for the horizontal polarization single input, one can find $P(D_{13}|2_H,0)=P(D_{13}|0,2_H)=1/2$, and all other $P(D_{ij}|2_H,0)=P(D_{ij}|0,2_H)=0$. Likewise, the vertical single input gives $P(D_{24}|2_V,0)=P(D_{24}|0,2_V)=1/2$, while all other $P(D_{ij}|2_V,0)=P(D_{ij}|0,2_V)=0$. Therefore, except for $D_{13}$ and $D_{24}$, (\ref{input}) in $Z$-basis becomes simplified as
%%%%%%%%%%%%%%%%%%%%%%%%%
\begin{eqnarray}
N(D_{ij})^{\mu,\mu}=\kappa_i\kappa_j\mu^2e^{-2\mu}P(D_{ij}|1_{H/V},1_{H/V}).
\label{z_out}
\end{eqnarray}
%%%%%%%%%%%%%%%%%%%%%%%%%
Considering $D_{13}$ and $D_{24}$ does not account for the BSA result, one can find that, in $Z$-basis, the linear optical BSA results with WCP are identical to those with ideal single-photon inputs. This result coincides with that the QBER in $Z$-basis (when both Alice and Bob choose $Z$-basis) using WPC can be $Q^{\mu,\mu}_{ZZ}=0$, and thus, is used for secret key distribution~\cite{lo12}. %Here, the subscript $RS$ stands for Alice and Bob choose $R$ and $S$-bases, respectively.

In $X$- and $Y$-basis, the difference between WPC and ideal single-photon states becomes visible. Since $P(D_{ij}|2_P,0)$ and $P(D_{ij}|0,2_Q)$ contribute to the BSM results, the QBERs in $X$- and $Y$-bases with the WCP cannot be lower than $Q^{\mu,\mu}_{XX},\,Q^{\mu,\mu}_{YY}\ge0.25$ whereas those with single-photon states can be $Q^{1,1}_{XX},\,Q^{1,1}_{YY}=0$. For the input polarization states in these bases, $|P\rangle=\frac{1}{\sqrt{2}}\left(|H\rangle+e^{i\theta_{p}}|V\rangle\right)$, the single count probability becomes $P(D_i|1_P,0)=P(D_i|0,1_Q)=1/4$ for all $D_i$. Therefore, from Eqs.~(\ref{P_11}) and (\ref{D_i}), we find
%%%%%%%%%%%%%%%%%%%%%%%%%
\begin{eqnarray}
P(D_{ij}|1_P,1_Q)=\frac{N(D_{ij})^{\mu,\mu}-N(D_{ij})^{\mu,0}-N(D_{ij})^{0,\mu}}{\frac{1}{4}N(D_i)N(D_j)}.\nonumber\\
\label{P_11_xy}
\end{eqnarray}
%%%%%%%%%%%%%%%%%%%%%%%%%
Equation~(\ref{P_11_xy}) implies that the coincidence probability $P(D_{ij})$, i.e., the BSM results, for ideal single-photon input states can be deduced from coincidence and single counts using WCP inputs. With the results, we can infer QBERs for ideal single-photon inputs of $Q_{RS}^{1,1}$ in $R$ and $S$ bases for Alice and Bob where $R,S\in\{X,Y\}$. Note also that all the experimental data $N(D_{ij})$, $N(D_i)$, and $N(D_j)$ in Eq.~(\ref{P_11_xy}) can be obtained from the ordinary decoy based MDI-QKD experiment.

%Moreover, QBER in $RS$ basis, where $R,S\in\{X,Y\}$, can be obtained as
%arbitrary basis consisted of $X$- and $Y$- basis, $Q^{1,1}_{RS}$ where $R,S\in\{X,Y\}$ can be also obtained as
%For clearance, we rewrite the coincidence probability, $P(D_{ij})$ into $P(\Psi^{\pm})$, which means the probability of Bell states projection according to detector relation as $\Psi^{+}=(D_{12},D_{34})$ and $\Psi^{-}=(D_{14},D_{23})$. %
%%%%%%%%%%%%%%%%%%%%%%%%%
%\begin{eqnarray}
%Q^{1,1}_{RS}=\frac{P(D_{ij}|1_{P},1_{Q})_{error}}{\sum P(D_{ij}|1_P,1_Q)},\label{Q_XX}
%\end{eqnarray}
%%%%%%%%%%%%%%%%%%%%%%%%%
%where $P(D_{ij}|1_{P},1_{Q})_{error}$ is the error proportion for BSM results, accordingly, $1_{P}$ and $1_{Q}$ are ones of polarization states in $R$- and $S$-basis. Note that Eq.~\eqref{P_11_xy} requires observing single and coincidence counts while blocking one of the inputs, however, it does not cause further experimental complexity since the vacuum decoy has already been widely utilized in most QKD system~\cite{yin16,liu19}.

%%%%%%%%%%%%%%%%%%%%%%%%%
\begin{figure}[t]
\centering\includegraphics[width=3in]{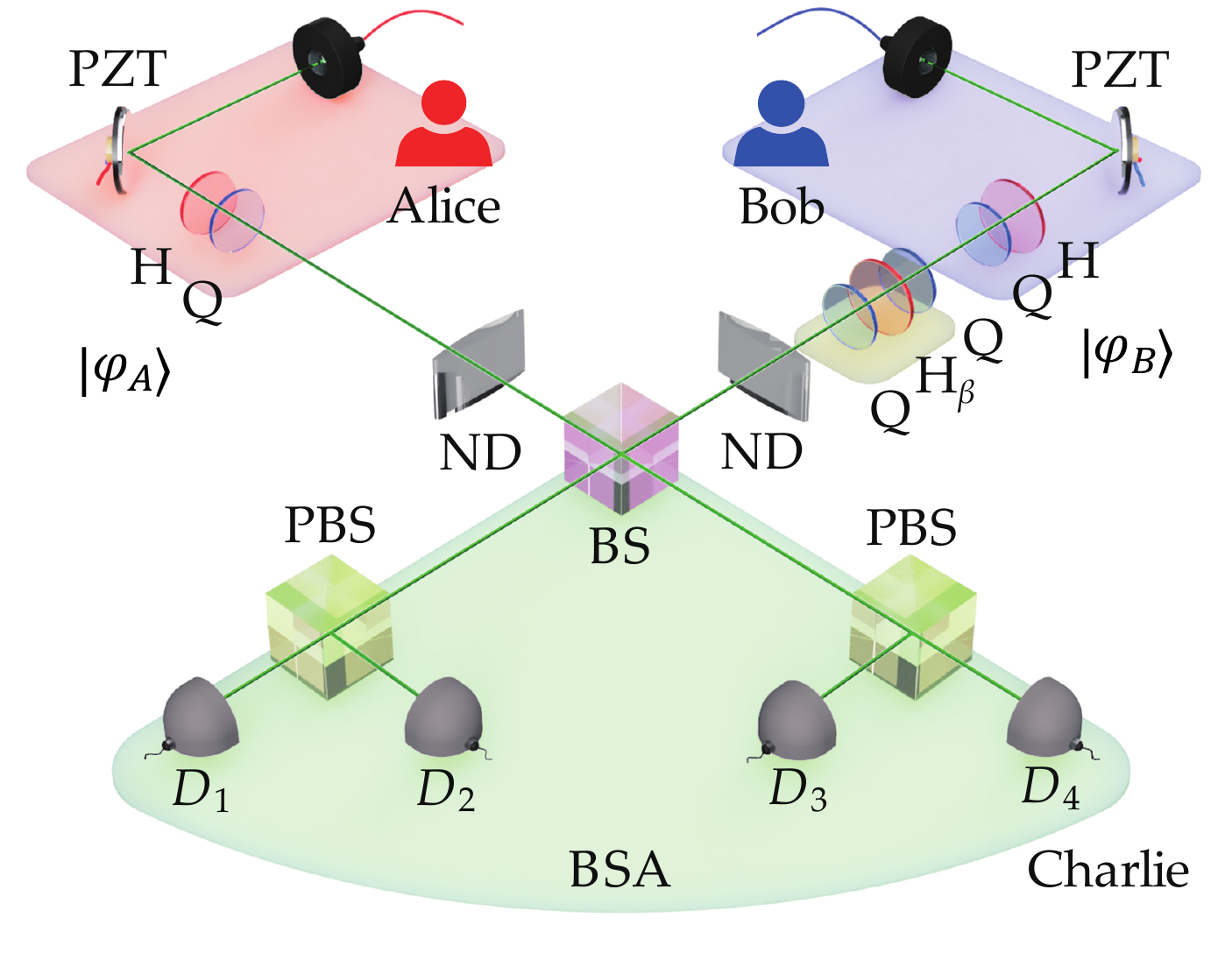} 
\caption{Experimental setup. $H$: half waveplate, $Q$: quarter waveplate, PZT: mirror with piezoelectric translator, ND: neutral density filters, BS: beamsplitter, PBS: polarizing beamsplitter, $D_{i}$ : single-photon detectors.}
\label{setup}
\end{figure}
%%%%%%%%%%%%%%%%%%%%%%%%%

Applying the above method to RFI-MDI-QKD using WCP finds more useful results in security analysis. In RFI-MDI-QKD, the security analysis requires the QBER in $Z$-basis, $Q^{1,1}_{ZZ}$ and the $C$ parameter given as~\cite{wang15}
%%%%%%%%%%%%%%%%%%%%%%%%%%%%%%%
%\begin{widetext}
\begin{eqnarray}
C&=&{\langle}X_AX_B{\rangle}^2+{\langle}X_AY_B{\rangle}^2+{\langle}Y_AX_B{\rangle}^2+{\langle}Y_AY_B{\rangle}^2\nonumber\\
&=&(1-2Q^{1,1}_{XX})^2+(1-2Q^{1,1}_{YY})^2+(1-2Q^{1,1}_{XY})^2\nonumber\\
&&+(1-2Q^{1,1}_{YX})^2.
\label{C-pameter}
\end{eqnarray}
%\end{widetext}
%%%%%%%%%%%%%%%%%%%%%%%%%%%%%%%
It is notable that both $Q^{1,1}_{ZZ}$ and $C$ are independent of the reference frame rotation, and thus, it does not require pre-shared reference frames between Alice and Bob. Note that $Q^{1,1}_{ZZ}=0$ and $C=2$ for RFI-MDI-QKD with ideal single-photon input states. While $Q^{1,1}_{ZZ}=Q^{\mu,\mu}_{ZZ}$ can be directly obtained using WCP, calculating $C$ requires $Q^{1,1}_{RS}$. Our method provides simple way to obtain $Q^{1,1}_{RS}$ using WCP. Comparing to the conventional way to calculate $C$ in RFI-MDI-QKD which requires accurate calibration of the average photon numbers of signal and decoy states, QBERs with different average photon number inputs, gains for different average photon number inputs, etc (see Eqs.~(6) and (7) of Ref.~\cite{wang15}), our method provides much more simple and robust mean to estimate $C$. Note that the recently developed RFI-MDI-QKD using fewer quantum states can also be implemented with this method in order to simplify the experimental implementation of RFI-MDI-QKD~\cite{lee20}. 

%Similar to $Q_{XX}$ of MDI-QKD, however, the $C$ parameter calculation using WCP includes inherent errors due to non-zero $P(D_{ij}|2_P,0)$ and $P(D_{ij}|0,2_Q)$ terms. In order to find the $C$ parameter for single-photon inputs, one needs to accurately calibrate the average photon numbers of signal and decoy states, QBER with different average photon number inputs, gain for different average photon number inputs, etc, and perform complicated numerical calculation, for example, see Eq.~(6)-(7) of Ref.~\cite{wang15}. The $C$-parameter can be much more easily calculated with Eq.~\ref{P_11_xy}. Note that the recently developed RFI-MDI-QKD using fewer quantum states can be also applicable~\cite{lee20}.

%%%%%%%%%%%%%%%%%%%%%%%%%
%\begin{eqnarray}
%&&P(D_{13}|2_H,0)=P(D_{13}|0,2_H)=\frac{1}{2},\nonumber\\
%&&P(D_{ij}|2_H,0)=P(D_{ij}|0,2_H)=0,~{\rm for~other}~D_{ij},\nonumber\\
%&&P(D_{24}|2_V,0)=P(D_{24}|0,2_V)=\frac{1}{2},\\
%&&P(D_{ij}|2_V,0)=P(D_{ij}|0,2_V)=0,~{\rm for~other}~D_{ij}.\nonumber
%\end{eqnarray}
%%%%%%%%%%%%%%%%%%%%%%%%%

\section{Experiment} 

%\noindent {\it Experiment.--} 

In order to compare the deduced BSM results using WCP and ideal single-photon inputs, we performed the experiments using attenuated laser pulses and photon pairs from spontaneous parametric down conversion (SPDC). The single-photon pairs at 1556~nm are generated by type-II SPDC using 10 mm periodically-polled KTP crystal pumped by femtosecond laser pulses. The WCP are obtained by attenuating femtosecond laser pulses. In order to make the spectral bandwidths identical, both light sources are flittered by the same interference filters with 3~nm bandwidth. %Superconducting nanowire single-photon detectors (SNSPDs) are utilized for efficient single-photon detection. 
Then, the optical pulses are sent to Alice and Bob who correspond to the transmitters of MDI-QKD, see Fig.~\ref{setup}.

%%%%%%%%%%%%%%%%%%%%%%%%%
\begin{figure}[t]
\centering\includegraphics[width=3.45in]{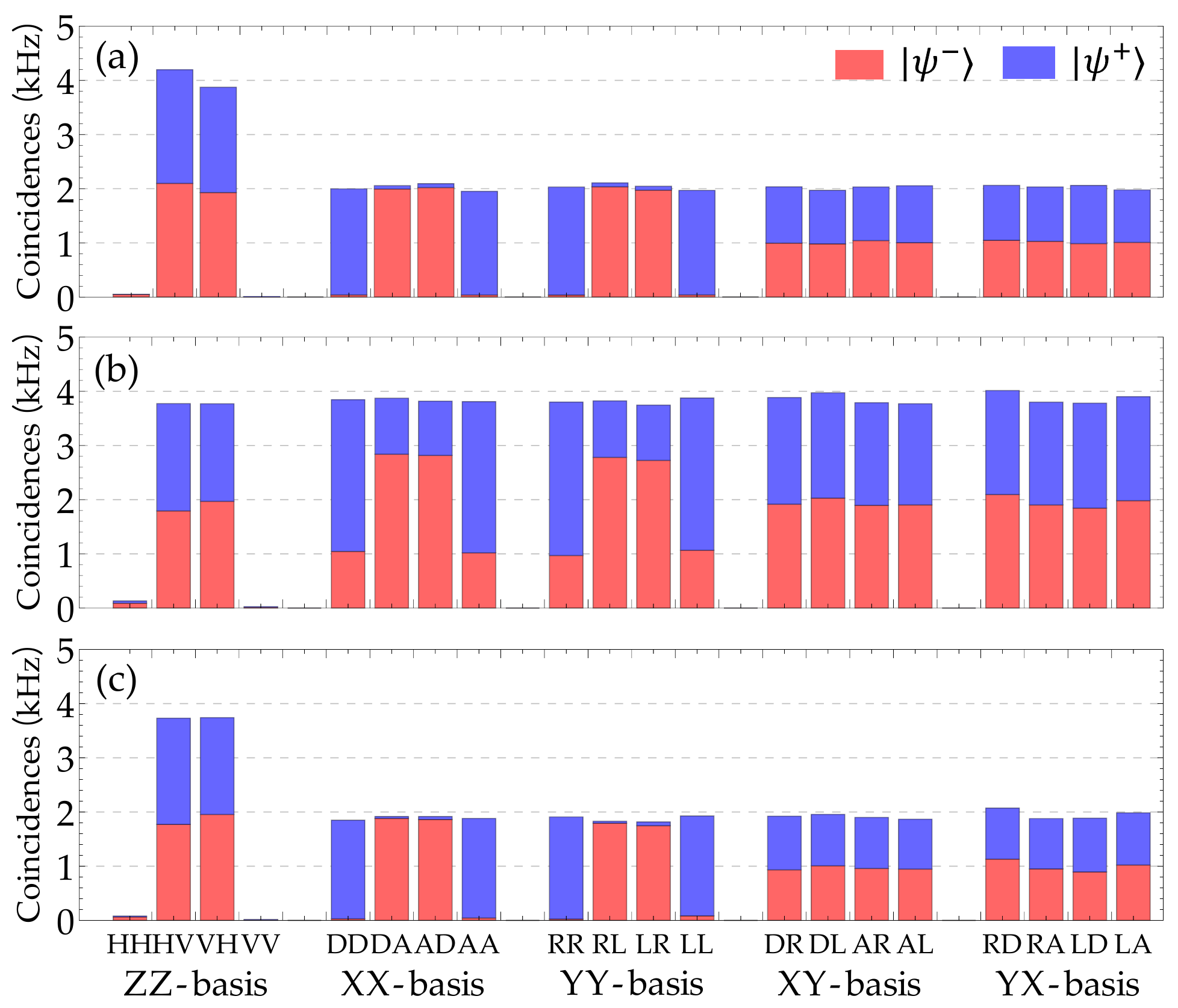} 
\caption{Bell state measurement results with (a) single-photon inputs from SPDC, (b) WCP with $\mu=0.25$, and (c) deduced single-photon inputs using WCP. Here, $\mu$ is determined at the outputs of Alice and Bob, and thus the effective average photon number at Charlie becomes $\mu\sim0.008$ after 15~dB of quantum channel loss.}
\label{BSM}
\end{figure}
%%%%%%%%%%%%%%%%%%%%%%%%% 

%%%%%%%%%%%%%%%%%%%%%%%%%%%%%
\begin{table}[b]
\centering\begin{tabular}{|c|c|c|c|}
\hline
Inputs                     & $ZZ$-basis        & $XX$-basis        & $YY$-basis        \\ \hline
~~$Q^{\rm SPDC}$~~         & ~$0.01\pm0.001$~  & ~$0.034\pm0.002$~ & ~$0.038\pm0.002$~ \\ \hline
$Q^{\mu,\mu}$             & $0.018\pm0.001$ & $0.269\pm0.004$ & $0.265\pm0.004$ \\ \hline
$Q^{1,1}$ & $0.014\pm0.002$ & $0.037\pm0.013$ & $0.030\pm0.013$ \\ \hline
\end{tabular}\label{qber}
\caption{QBERs in MDI-QKD scenario for different bases with various inputs. $Q^{\rm SPDC}$: SPDC, $Q^{\mu,\mu}$: WCP, $Q^{1,1}$: deduced single-photon inputs using WCP.}
\end{table}
%%%%%%%%%%%%%%%%%%%%%%%%%%%%%

In order to erase the first-order interference of WCP, Alice and Bob employ mirrors attached with piezoelectric translators (PZT)~\cite{kim13,kim14}. The PZTs are independently modulated during the experiment. The polarization states of $|\varphi\rangle_A$ and $|\varphi\rangle_B$ are encoded using half-, and quarter-waveplates (H, Q), then the optical pulses are sent to Charlie who performs BSM using a linear optical BSA. The 15~dB neutral density filters (ND) at the optical paths simulate the quantum channel loss in the QKD communication. In order to obtain $P(D_{ij}|1_P,1_Q)$ using WCP, the BSM was performed when both Alice and Bob send optical pulses, $N(D_{ij})^{\mu,\mu}$, and only one of Alice and Bob sends an optical pulse, $N(D_{ij})^{\mu,0}, N(D_{ij})^{0,\mu},$ and $N(D)_i$. Note that, in the MDI-QKD scenario, the later corresponds to the case when one of the transmitters sends signal state while the other sends vacuum decoy state.

Figure~\ref{BSM} shows the BSM results with (a) single-photon inputs from SPDC, (b) WCP with $\mu=0.25$, and (c) deduced single-photon inputs using WCP. Here, $\mu$ is determined at the output of Alice and Bob, and thus, the effective average photon number at Charlie becomes $\mu\sim0.008$ after 15~dB of quantum channel loss. It shows that the BSM results for $ZZ$-basis inputs are all similar among different optical inputs. On the other hands, the BSM results for other bases present the difference. The BSM results with WCP are clearly different from those with SPDC single-photon inputs. However, one can obtain very similar results with the SPDC single-photon inputs by deducing single-photon inputs using WCP. This can be quantified by QBERs in MDI-QKD scenario, see TABLE~I. For $ZZ$-basis, QBERs with all inputs are close to $Q_{ZZ}=0$. For $XX$, and $YY$-bases, however, QBER with WCP is much higher than those with SPDC inputs. Note that the intrinsic QBER limit with WCP in these bases is $Q_{XX}=Q_{YY}=0.25$. By applying our method to deduce the single-photon inputs, QBERs become as low as those with SPDC inputs. Note, however, lowering QBERs does not mean that we can utilize these bases for secret key generation. It happens as a result of statistical treatment and does not effective on the individual events.

%Note that QBERs in $ZZ$ and $XX(YY)$-bases are calculated as
%%%%%%%%%%%%%%%%%%%%%%%%%%%%%%%
%\begin{eqnarray}
%Q_{ZZ}=\frac{C_{HH}+C_{VV}}{C_{HH}+C_{HV}+C_{VH}+C_{VV}},\\
%Q_{XX}=\frac{}{}
%\label{QBER}
%\end{eqnarray}
%%%%%%%%%%%%%%%%%%%%%%%%%%%%%%%

%It shows that the BSM results for $ZZ$-basis inputs are the same regardless of the optical inputs. The data can be quantified by QBER in MDI-QKD. With Fig.~\ref{BSM}, we can obtain similar $Q_{ZZ}$ regardless of the input states, i.e., $Q_{ZZ}^{\rm SPDC}=0.01\pm0.001$, $Q_{ZZ}^{\mu,\mu}=0.018\pm0.001$, and $Q_{ZZ}^{1,1}=0.014\pm0.002$, respectively.

%The difference of the input optical states becomes visible for $XX$ and $YY$-bases. For single-photon inputs from SPDC, the QBERs are calculated as $Q_{XX}^{\rm SPDC}=0.034\pm0.002$, and $Q_{YY}^{\rm SPDC}=0.038\pm0.002$. On the other hands, the inputs of WCP provides $Q_{XX}^{\mu,\mu}=0.269\pm0.004$, and $Q_{YY}^{\mu,\mu}=0.265\pm0.004$. These results coincide with the non-zero intrinsic MDI-QKD QBER bound of $Q=0.25$ in $XX$- and $YY$-bases for WCP. Fig.3~\ref{BSM} (c) clearly shows that the deduced BSM results of single-photon inputs using WCP are similar those of SPDC photons. The inferred QBERs of $Q_{XX}^{1,1}=0.037\pm0.013$ and $Q_{YY}^{1,1}=0.030\pm0.013$ show that we can indeed obtain the QBERs for single-photon inputs using WCP.

%%%%%%%%%%%%%%%%%%%%%%%%%
\begin{figure}[t!]
\centering\includegraphics[width=3.45in]{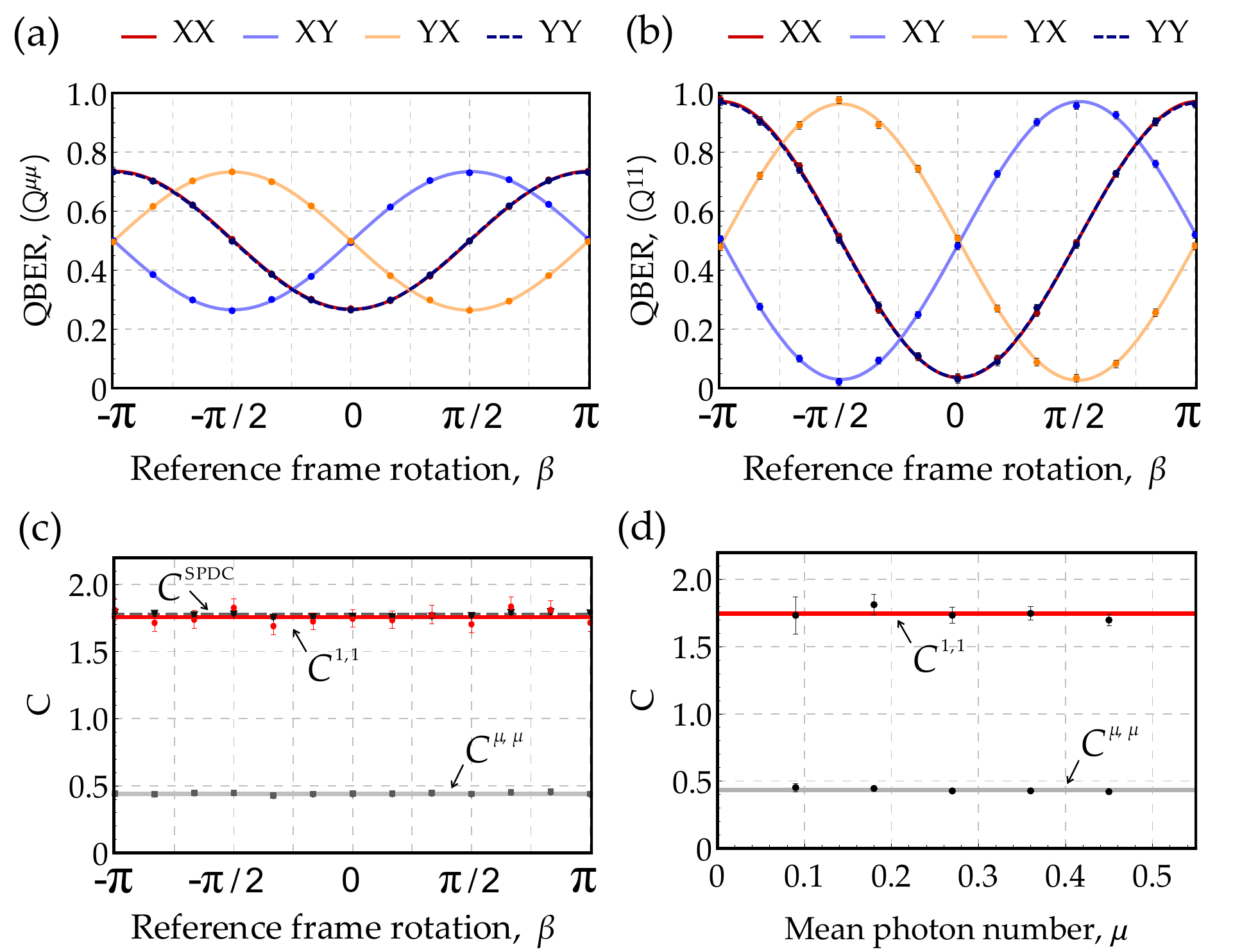} 
\caption{Experimental results of RFI-MDI-QKD. QBERs for (a) WCP with $\mu=0.25$, and (b) deduced single-photon inputs using WCP. $C$ is presented with respect to (c) reference frame rotation $\beta$, and (b) mean photon number $\mu$. Error bars are obtained by repeating 1,000 times of Monte Carlo simulation by from the experiment data.}
\label{rfi}
\end{figure}
%%%%%%%%%%%%%%%%%%%%%%%%% 

In order to investigate the effectiveness of our method in RFI-MDI-QKD, we have performed the protocol with respect to reference frame rotation $\beta$ at Bob's channel. As shown in Fig.~\ref{setup}, the reference frame rotation was implemented by rotating a half-waveplate (${\rm H}_\beta$) which is located between two quarter waveplates at $45^\circ$~\cite{yoon19}. Figure~\ref{rfi} shows QBERs for (a) WCP and (b) deduced single-photon inputs using WCP with respect to $\beta$. The $\beta$ independent $Q_{ZZ} < 0.02$ is not presented. By deducing the single-photon inputs, the visibility of the sinusoidal oscillation increases from $V^{\mu,\mu}=0.47\pm0.002$ to $V^{1,1}=0.94\pm0.01$. Figure~\ref{rfi} (c) presents estimated $C$ for various inputs with respect to $\beta$. While all $C$ are invariant under the reference frame rotation, it clearly shows $C^{1,1}$ for deduced single-photon inputs using WCP is similar with $C^{\rm SPDC}$ for single-photon inputs using SPDC. The estimated $C$ are $C^{\mu,\mu}=0.48\pm0.002$, $C^{1,1}=1.75\pm0.014$, and $C^{\rm SPDC}=1.77\pm0.004$, respectively. Different mean photon numbers for WCP does not provide much difference in calculating $C^{\mu,\mu}$ and $C^{1,1}$, see Fig.~\ref{rfi} (d). These results clearly shows that $C$ can be obtained with our method which does not require precise calibration of QBERs and gains with optical pulses with different average photon numbers.

\section{Conclusion}

%\noindent {\it Conclusion.--} 

To summarize, we have proposed and experimentally verified a method to deduce BSA for single-photon inputs using WCP. We have also applied the method to RFI-MDI-QKD and verified the effectiveness in estimating the security associated parameter $C$. We note that applying our method to MDI- and RFI-MDI-QKD in more realistic experimental conditions including the finite key size analysis would be necessary for future work. We also remark it would be an interesting research direction to extend our methods to larger linear optical circuits with a large number of inputs such as GHZ state measurement.%~\cite{navarrete18,zhang20}.

\begin{acknowledgements}
This work was supported by the NRF programs (2019M3E4A1079777, 2019R1A2C2006381, 2019M3E4A107866011), the IITP programs (2020-0-00947, 2020-0-00972), and the KIST research program (2E30620).
\end{acknowledgements}

%\section*{Conflict of interest}
%The authors declare that they have no conflict of interest.

\end{document}